\documentclass [12pt]{article}

\usepackage {graphicx}
\usepackage{cite}

\begin{document}

\begin{center}
 \large{Chaotic Dynamics of a Nonlinear Ring Cavity \\
Driven by an External Multi-frequency Signal}
\end{center}

\begin{center}
{A.A. Balyakin, N.M. Ryskin \\
\textit{Saratov State University, Department of Nonlinear
Physics}\\
\textit{83, Astrakhanskaya str., Saratov, Russia,410012}\\
\textit{e-mail: BalyakinAA@info.sgu.ru}}
\end{center}

Keywords: Chaos; Nonlinear ring cavity; Optical bistability; Ikeda
map; Multi-frequency driving

PACS: 42.65.Sf; 47.52+j; 05.45.-a; 0545.Jn

\section{Introduction}

A nonlinear ring cavity driven by an external signal is among the
most popular models of nonlinear dynamics \cite{Neimark, Landa}. A
special attention to this system is paid in non-linear optics
where it was considered for the first time by Ikeda and co-authors
\cite{Neimark, Landa, Ikeda1, Ikeda2, Ikeda3}. Many interesting
non-linear phenomena such as bistability, transitions to chaos,
and pattern formation have been studied both theoretically and
experimentally (see for details \cite {Gibbs, Rosanov,Moloney}).
Those researches are of great interest because of prospects to
apply bistable cavities as logical elements in optical computers
and in cryptography \cite{Gibbs,Rosanov,Wang,Alvarez}. We note that similar
systems can be also produced not only in optics, but also in
microwave diapason.

Schematic drawing of studied device is presented in Fig.~1. A
nonlinear dielectric with Kerr-type nonlinearity is placed inside
a ring optical cavity. The cavity is excited by an input signal
$E_0 \left( t \right)$ from an external laser source. However, we
do not restrict ourselves to the nonlinear optical system. Our
analysis is valid for any type of waves propagating in a nonlinear
medium with cubic phase nonlinearity when output signal is partly
recirculated to the input through a feedback loop.

For the case of harmonic input signal the behavior of the system
has been studied thoroughly \cite{Neimark, Landa, Ikeda1, Ikeda2,
Ikeda3,Gibbs, Rosanov,Moloney,Chesnokov}. Under several
assumptions (plane wave approximation, neglecting dispersion and
losses in the nonlinear media, etc.) a simple return map can be
derived to describe the dynamics of complex amplitude of the
signal

\begin{equation}
\label{eq1}
A_{n + 1} = A_0 + \rho A_n \exp \left( {i\left| {A_n } \right|^2 + i\varphi
} \right).
\end{equation}

Here $A_n $ is the signal amplitude on $n$-th spreading through
the cavity, $A_0 $ is a constant amplitude of the input signal,
$\rho $ is an amount of feedback ($0 < \rho < 1)$, and $\varphi $
is a total phase shift of the signal for a single spreading
through the feedback loop. The map (\ref{eq1}) is known as
\textit{Ikeda map}. For sufficiently high input power the
steady-state single frequency oscillations become unstable and
periodic pulsations arise with the period equal to doubled delay
time \cite{Neimark, Landa, Ikeda1, Ikeda2, Ikeda3,Gibbs,
Rosanov,Moloney}. This phenomenon is called Ikeda instability. The
subsequent increase of the input signal leads to transition to
chaos by Feigenbaum scenario and then to formation of Ikeda
attractor \cite{Ikeda1, Ikeda2} that corresponds to the
fully-developed, strongly irregular chaotic motion.

In this paper, we investigate nonlinear dynamics of this system when the
input signal is multi-frequency, containing several discrete spectral
components $\omega _1 ,\omega _2 ,..,\omega _n $:

\begin{equation}
\label{eq2}
E_0 = \sum\limits_{j = 1}^n {A_{0j} \exp \left( {i\omega _j t} \right)} +
\mbox{c.c.},
\end{equation}

\noindent where $A_{0j} $ are complex constant amplitudes, c.c.
denotes complex conjugate. For the sake of simplicity we suppose
the frequencies $\omega _j $ to be non-resonant, i.e. their ratios
are irrational numbers, and restrict ourselves to double-frequency
driving, $n = 2$. Note that similar problem (for $n = 2$ or 3)
have been already considered in \cite{Pliszka}. However, in cited
work the main concern was with the steady-state characteristics of
the device in the case of symmetrical pumping (i.e. equal input
intensities for both frequencies). In contrast to  \cite{Pliszka}
we will focus on case when the pumping is strongly asymmetric,
treating the second signal as a small control action. Our aim is
to demonstrate that varying the intensity of the second signal
opens a possibility to effective control of the dynamics of the
primary signal. We also will concentrate on the transitions to
chaos and on peculiarities of the chaotic oscillations. Some
preliminary results presented in  \cite{Balyakin1, Balyakin2} show
that the chaotic dynamics can be much more diverse than for the
single-frequency driving.

\begin{figure}[h]
\centerline{\includegraphics{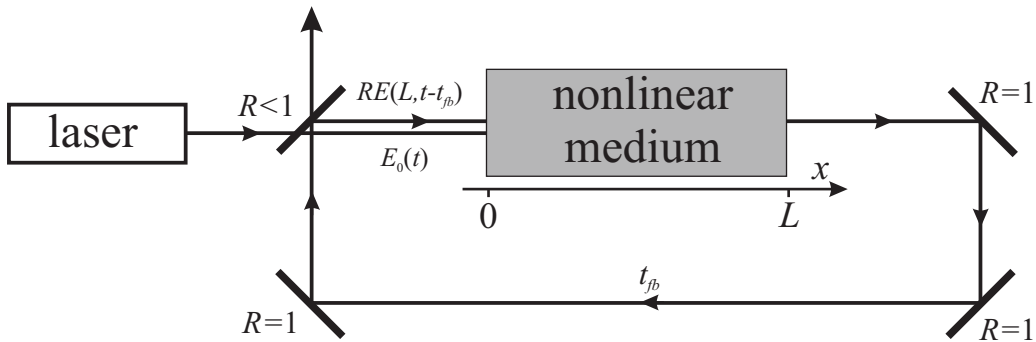}} \label{fig1} \caption
{\small Schematic drawing of a nonlinear ring cavity under
external driving}
\end{figure}

The paper is organized as follows. In Sec.~\ref{sec:derivation},
the derivation of the coupled Ikeda maps is described in details.
Although the equations are obtained for a general case of an
arbitrary number of input frequencies, further the
double-frequency driving is considered. The stability analysis of
obtained equations is performed in Sec.~\ref{sec:stat}. Results of
numerical simulations of non-stationary oscillations are presented
in Sec.~\ref{sec:results}. Various types of transitions to chaos
are discovered while the input signal power is increased.
Sec.~\ref{sec:concluding} contains some concluding remarks.

\section{Derivation of coupled Ikeda maps}
\label{sec:derivation}

To describe the propagation of a multi-frequency wave packet in a
nonlinear medium we use the Hamiltonian formalism that is one of
the most powerful tools in the theory on non-linear wave
interaction (for review see \cite{Zakharov}). We assume that only
one type of waves should be considered, with dispersion relation
$\omega = \omega \left( k \right)$ that does not admit three-wave
resonant interactions. Thus, first-order nontrivial nonlinear
processes are caused by four-wave interactions, and Hamiltonian of
the system can be written as follows \cite{Zakharov}:

\begin{equation}
\label{eq3}
H = H_0 + H_{int} + ...,
\end{equation}

\begin{equation}
\label{eq4}
H_0 = \int {\omega \left( k \right)a_k a_k^ * dk} ,
\end{equation}

\begin{equation}
\label{eq5}
H_{int} = \frac{1}{2}\int {T_{k_1 k_2 k_3 k_4 } a_{k_1 }^ * a_{k_2 }^ *
a_{k_3 } a_{k_4 } \delta \left( {k_1 + k_2 - k_3 - k_4 } \right)dk_1 dk_2
dk_3 dk_4 } .
\end{equation}

Here $H_0 $ is a linear part of Hamiltonian, $H_{int} $ describes
nonlinear interactions. In (\ref{eq3})-(\ref{eq5}) $a_k $ is a
Fourier amplitude of the wave with wavenumber $k$, $\delta $ is
the Dirac delta-function, $T_{k_1 k_2 k_3 k_4 } $ is a four-wave
interaction matrix element (see \cite{Zakharov} for more details).

For the input signal (\ref{eq2}) we suppose the field to be presented as a
multi-frequency wave packet consisting of several discrete spectral
components

\begin{equation}
\label{eq6}
E = \sum\limits_{j = 1}^n {A_j \left( {x,t} \right)\exp \left( {i\omega _j t
- ik_j x} \right) + \mbox{c.c.},}
\end{equation}

\noindent
with frequencies $\omega _j $ and wavenumbers $k_j $ being connected with
each other by dispersion relation $\omega _j = \omega \left( {k_j } \right)$
and $A_j $ being the slow varying (in comparison with complex exponents)
envelope amplitudes. If all the frequencies are non-resonant, i.e. none of
the conditions

\[
\sum\limits_j {N_j \omega _j } = 0
\]

\noindent holds true for integer $N_j $, one can show that
propagation of the wavepacket (\ref{eq6}) is governed by the
system of coupled non-linear Schr\"{o}dinger equations
\cite{Ryskin}

\begin{equation}
\label{eq7}
i\left( {\frac{\partial A_j }{\partial t} + V_g^j \frac{\partial A_j
}{\partial x}} \right) + \frac{{\omega }''_j }{2}\frac{\partial ^2A_j
}{\partial x^2} + \left( {\beta _{jj} \left| {A_j } \right|^2 +
\sum\limits_{j \ne i} {\beta _{ji} } \left| {A_i } \right|^2} \right)A_j =
0.
\end{equation}

Here $V_g^j = \left. {\frac{d\omega }{dk}} \right|_{k = k_j } $
are the group velocities, ${\omega }''_j = \left. {\frac{d^2\omega
}{dk^2}} \right|_{k = k_j } $ are the parameters of group velocity
dispersion, and coefficients $\beta $  responsible for non-linear
interaction between waves with different components of the wave
packet are defined as follows:

\begin{equation}
\label{eq8}
\beta _{jj} = - 2T_{k_j k_j k_j k_j } ,
\quad
\beta _{ji} = - \left( {T_{k_j k_i k_j k_i } + T_{k_j k_i k_i k_j } + T_{k_i
k_j k_j k_i } + T_{k_i k_j k_i k_j } } \right).
\end{equation}

Further we shall neglect the effects of group velocities dispersion
(${\omega }''_j = 0$, $V_g^j \equiv V_g )$. Hence, the equations (\ref{eq7})
transforms into

\begin{equation}
\label{eq9}
i\left( {\frac{\partial A_j }{\partial t} + V_g \frac{\partial A_j
}{\partial x}} \right) + \left( {\beta _{jj} \left| {A_j } \right|^2 +
\sum\limits_{j \ne i} {\beta _{ji} } \left| {A_i } \right|^2} \right)A_j =
0.
\end{equation}

One can easily find the solution of (\ref{eq9}) in the form

\begin{equation}
\label{eq10}
A_j (x,t) = A_j \left( {0,t - \frac{x}{V_g }} \right)\exp \left\{
{\frac{ix}{V_g }\left[ {\beta _{jj} \left| {A_j \left( {0,t - \frac{x}{V_g
}} \right)} \right|^2 + \sum\limits_{j \ne i} {\beta _{ji} } \left| {A_i
\left( {0,t - \frac{x}{V_g }} \right)} \right|^2} \right]} \right\}.
\end{equation}

Since the total field at the input of the non-linear medium (at $x = 0)$ is
composed by the external driving signal and the signal traveled through the
resonator (see Fig.~1) the boundary condition is the following

\begin{equation}
\label{eq11}
E\left( {0,t} \right) = E_0 + RE\left( {L,t - t_{fb} } \right).
\end{equation}

Here $t_{fb} $ is the time of signal propagation in the feedback loop, $L$
is the length of the non-linear medium, and $R$ is (generally complex)
feedback parameter which is assumed to be the same for all frequencies. For
the case of an optical ring cavity shown in Fig.~1 $R$ is the reflectivity
of the output mirror (other three mirrors are assumed to have $R = 1)$.

Substituting (\ref{eq2}), (\ref{eq6}), and (\ref{eq10}) into
(\ref{eq11}) we obtain a system of delayed equations (cf.
\cite{Ikeda3,Gibbs,Rosanov,Moloney})

\begin{equation}
\label{eq12}
A_j \left( {t + \tau } \right) = A_{0j} + \rho A_j \left( t \right)\exp
\left[ {i\left( {\frac{\beta _{jj} L}{V_g }\left| {A_j \left( t \right)}
\right|^2 + \sum\limits_{i \ne j} {\frac{\beta _{ji} L}{V_g }\left| {A_i
\left( t \right)} \right|^2} + \varphi _j } \right)} \right],
\end{equation}

\noindent
where $\tau = t_{fb} + l \mathord{\left/ {\vphantom {l {V_g }}} \right.
\kern-\nulldelimiterspace} {V_g }$ is the time delay, $\rho = \left| R
\right|$, $\varphi _j = \mbox{Arg}\left( R \right) - \omega _j \tau $ are
total phase shifts for a single spreading through the feedback loop, that in
general are different for different spectral components. Note that the phase
space of the system of time delayed equations (or functional maps) (\ref{eq12}) is
infinite-dimensional. Our next step is to turn to a more simple system of
coupled return maps. Examining the output signal dynamics in discrete
moments of time $t_n = n\tau $ yields a system of coupled Ikeda maps

\begin{equation}
\label{eq13}
A_j^{n + 1} = A_{0j} + \rho A_j^n \exp \left[ {i\left( {\frac{\beta _{jj}
L}{V_g }\left| {A_j^n } \right|^2 + \sum\limits_{i \ne j} {\frac{\beta _{ji}
L}{V_g }\left| {A_i^n } \right|^2} + \varphi _j } \right)} \right],
\end{equation}

\noindent
where $A_j^n = A_j \left( {t_n } \right)$.

We restrict ourselves now to the investigation of two-frequency external
signal. In that case we obtain from (\ref{eq13})

\begin{equation}
\label{eq14}
\begin{array}{l}
 A_1^{n + 1} = A_{01} + \rho A_1^n \exp \left[ {i\left( {\left| {A_1^n }
\right|^2\frac{\beta _{11} L}{V_g } + \left| {A_2^n } \right|^2\frac{\beta
_{12} L}{V_g } + \varphi _1 } \right)} \right], \\
 A_2^{n + 1} = A_{02} + \rho A_2^n \exp \left[ {i\left( {\left| {A_2^n }
\right|^2\frac{\beta _{22} L}{V_g } + \left| {A_1^n } \right|^2\frac{\beta
_{21} L}{V_g } + \varphi _2 } \right)} \right]. \\
 \end{array}
\end{equation}

Let us suppose all four-wave interaction matrix elements to be equal:
$\;T_{k_1 k_2 k_3 k_4 } = T$. Thus, equations (\ref{eq8}) give $\beta _{jj} \equiv
\beta = - 2T$, $\beta _{ji} = - 4T = 2\beta $. Finally, to minimize the
number of significant parameters we make in (\ref{eq14}) the following change of
variables: $A_j \to {A_j } \mathord{\left/ {\vphantom {{A_j } {\sqrt {\beta
L / V_g } }}} \right. \kern-\nulldelimiterspace} {\sqrt {\beta L / V_g } }$,
$A_{0j} \to {A_{0j} } \mathord{\left/ {\vphantom {{A_{0j} } {\sqrt {\beta L
/ V_g } }}} \right. \kern-\nulldelimiterspace} {\sqrt {\beta L / V_g } }$.
As a result we obtain

\begin{equation}
\label{eq15}
\begin{array}{l}
 A_1^{n + 1} = A_{01} + \rho A_1^n \exp \left[ {i\left( {\left| {A_1^n }
\right|^2 + 2\left| {A_2^n } \right|^2 + \varphi _1 } \right)} \right], \\
 A_2^{n + 1} = A_{02} + \rho A_2^n \exp \left[ {i\left( {\left| {A_2^n }
\right|^2 + 2\left| {A_1^n } \right|^2 + \varphi _2 } \right)} \right]. \\
 \end{array}
\end{equation}

Since the variables $A_j^n $ are complex, the phase space of this system is
four-dimensional. Note that in the absence of the second signal, $A_{02} =
0$, (\ref{eq15}) can be reduced to the single Ikeda map (\ref{eq1}).

We recall that coupled return maps (\ref{eq15}) derived in
\ref{sec:derivation} performs simple model (we used plane wave
approximation, neglected dispersion and losses in the nonlinear
media, etc.). Exact solution of coupled Sch\"{o}dinger equations
(\ref{eq7}) is also possible and will be presented hereafter. We
will not discuss here the difference between two approaches
either. Some remarks concerning differential equations with
time-delayed feedback can be found in, e.g.
\cite{Ikeda4,Hong,Hong2,Shahverdiev}

\section{Stationary regimes and stability analysis}
\label{sec:stat}

As \textit{stationary regimes} we name henceforth the oscillation
when signal amplitudes are time-independent. That corresponds to
stable points of coupled Ikeda maps $A_j^n = A_j^0 =
\mbox{const}$. For stationary intensities $I_j = \left| {A_j^0 }
\right|^2$ we have

\begin{equation}
\label{eq16} I_1 = \frac{I_{01} }{1 + \rho ^2 - 2\rho \cos \Phi _1
},
\end{equation}

\begin{equation}
\label{eq17} I_2 = \frac{I_{02} }{1 + \rho ^2 - 2\rho \cos \Phi _2
},
\end{equation}

\noindent here $I_{0j} = \left| {A_{0j} } \right|^2$ --
intensities of outcoming signals; $\Phi _i = I_i + 2I_j + \varphi
_i $, $i,j = 1,2$. Equations (\ref{eq16}), (\ref{eq17}) are
transcendental and can be solved only numerically. Fig. 2
represents the dependence $I_1 \left( {I_{01} } \right)$ --- gain
characteristics.

Dashed lines correspond to ${I_{01} } \mathord{\left/ {\vphantom
{{I_{01} } {\left( {1\pm \rho } \right)}}} \right.
\kern-\nulldelimiterspace} {\left( {1\pm \rho } \right)}^2$, that
is the limits of $I_{1,2} $:

\begin{equation}
\label{eq18} \frac{I_{0j} }{\left( {1 + \rho } \right)^2} < I_j <
\frac{I_{0j} }{\left( {1 - \rho } \right)^2}.
\end{equation}

Fig.2a corresponds to the case when the second signal is absent,
$I_{02} = 0$, and is analogous to well-known gain characteristics
of nonlinear ring cavity presented in, e.g., \cite{Rosanov}. That
diagram evidently illustrates the phenomena of multi-stability and
hysteresis, peculiar to discussed system. The branches with
negative incidence comply to unstable states. Arrows denote shifts
from one stable branch of gain characteristics to another.

\begin{figure}[h]
\leavevmode \centering
\includegraphics[width=1.0\columnwidth]{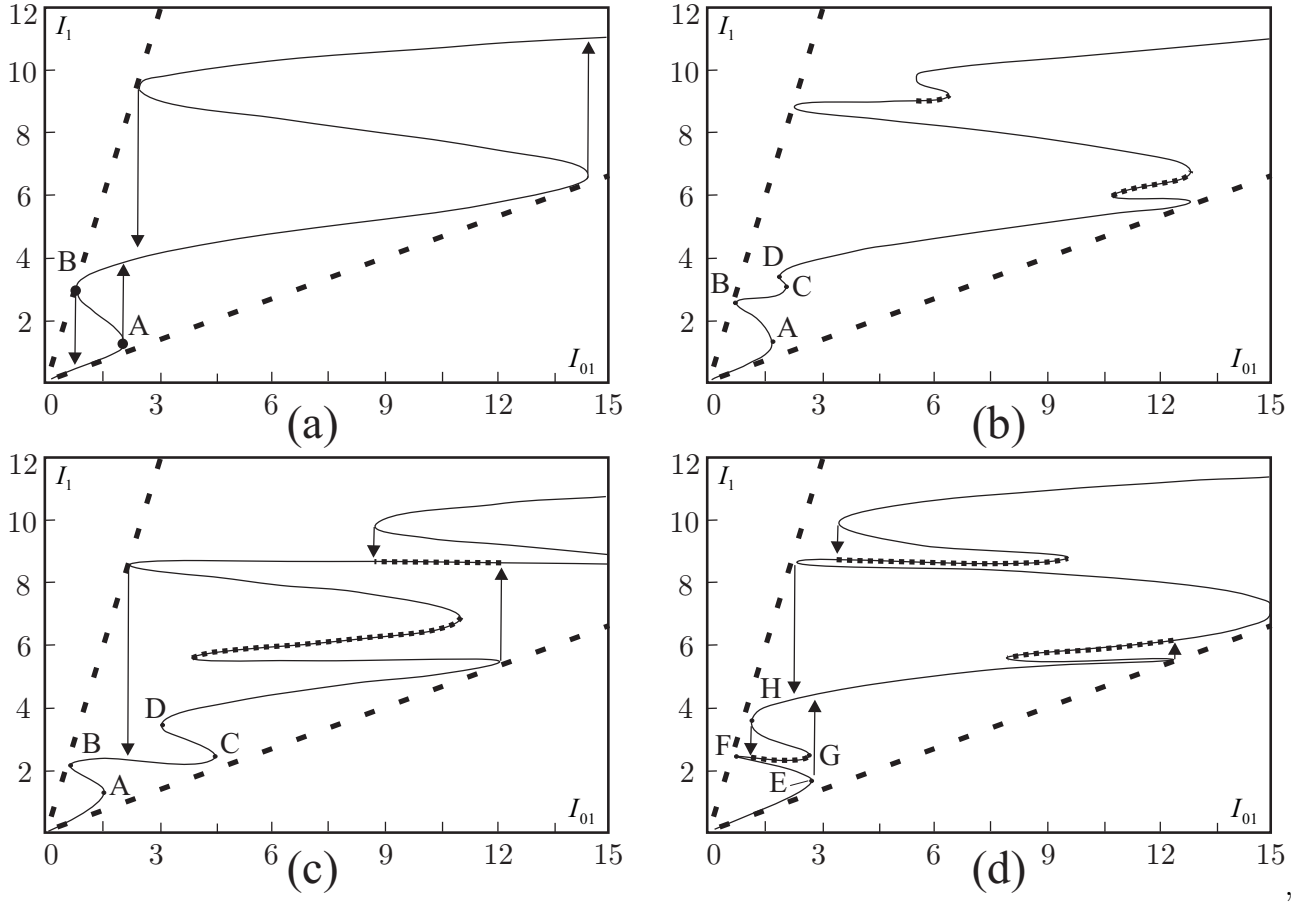},
\label{fig2} \caption{\small Gain characteristics for coupled
Ikeda maps. $\varphi_1$=$\pi$, $A_{02}$=0 (a), $A_{02}$=0.4 (b),
$A_{02}$=0.6 (c), $A_{02}$=0.6, $\varphi_1$=3$\pi$/4  (d)}
\end{figure}

The increase of the second signal intensity leads to the
complification of $I_1 \left( {I_{01} } \right)$, number of
branches also increases (Fig.2c). We note, that there exist stable
part of gain characteristics , that can not be arrived by smooth
varying of $I_{01} $ (they are marked with dots). To obtain these
regimes one should either provide very precise starting conditions
or move in parameter space in very particular direction. Thus
nonstationary dynamics of the system is expected to be much more
diverse comparatively to single Ikeda map. We emphasize that in
all cases performed in Fig.2 the intensity of the second component
remains rather small sofar it can be treated as weak control
action ($I_{02} < 0.36$, compare with $I_{01} )$.

Now let us undertake the stability analysis of stationary
solutions. Linearizing the maps in the vicinity of a stable point
after simple but bulky transformations we derived characteristic
equation:

\begin{equation}
\label{eq19}
\begin{array}{r}
 \left[ {\mu ^2 - 2\mu \rho \left( {\cos \Phi _1 - I_1 \sin \Phi _1 }
\right) + \rho ^2} \right]\left[ {\mu ^2 - 2\mu \rho \left( {\cos
\Phi _2 -
I_2 \sin \Phi _2 } \right) + \rho ^2} \right] = \\
 = 16\mu ^2\rho ^2I_1 I_2 \sin \Phi _1 \sin \Phi _2 . \\
 \end{array}
\end{equation}

\noindent here $\mu $ --- multiplicator. We note that terms in
brackets in the left part of (\ref{eq19}) coincide with
characteristic equations for single Ikeda map (\ref{eq1}).
Consequently we interpret the right-hand term as a coupling
parameter.

\begin{figure}[t]
\leavevmode \centering
\includegraphics[width=0.7\columnwidth]{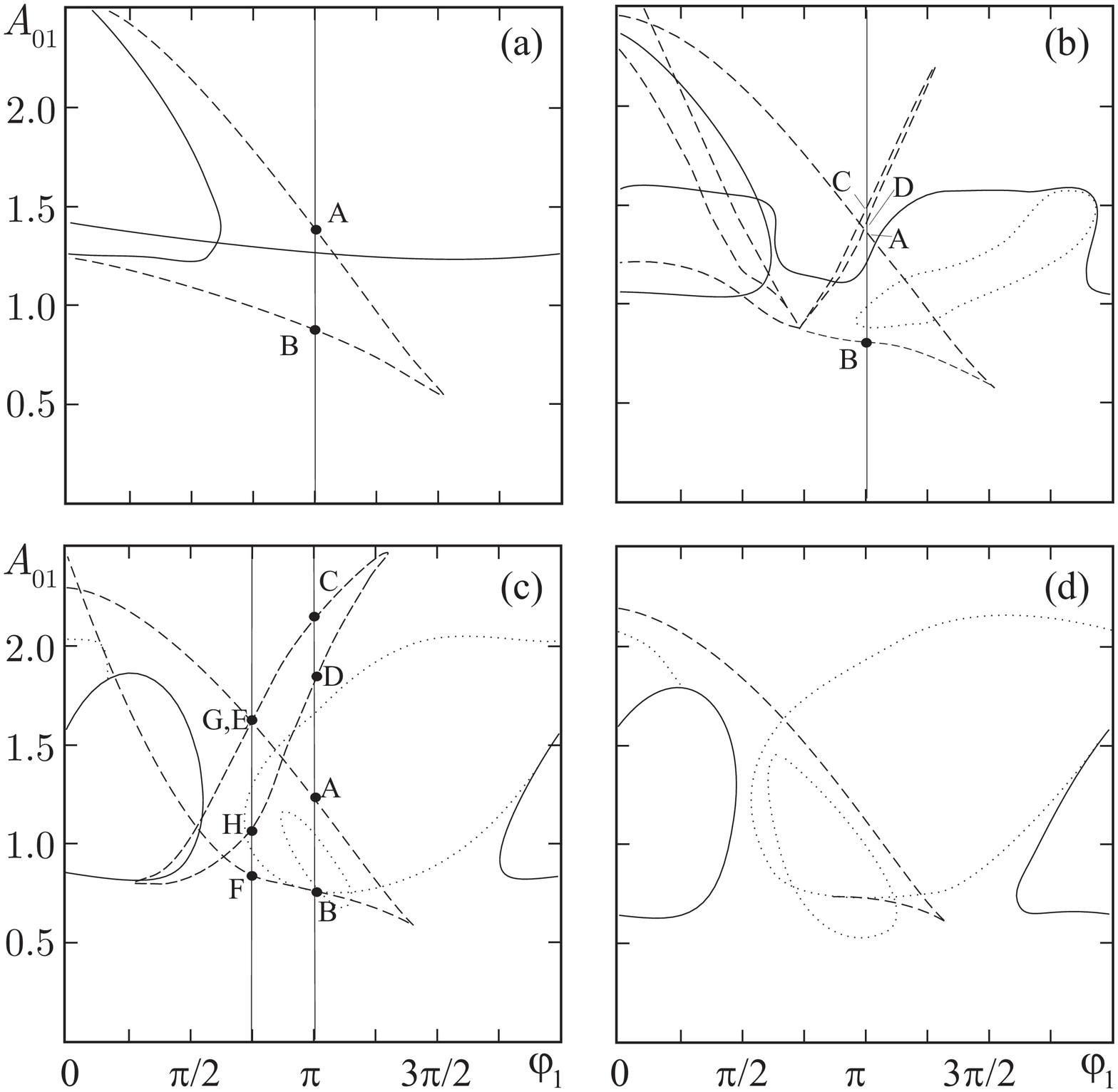},
\label{fig3} \caption{\small Lines of different bifurcations on a
parameter plane: first doubling bifurcation (solid lines), tangent
bifurcation (dash lines), Neimark bifurcation(dotted
lines).Parameters: $\rho = 0.5$, $\varphi _2 = 0$, and $A_{02} =
0.0$ (a), 0.4 (b), 0.6 (c), 0.8 (d). Thin lines correspond to
direction along which gain characteristics on Fig.2 were built }
\end{figure}

The loss of stability can happen in a various ways (see for
details, e.g., \cite{Ott}): as a result of doubling period
bifurcation ($\mu = - 1)$, tangent bifurcation ($\mu = 1)$, as
well as Neimark bifurcation (the last corresponds to two comlex
conjugate multiplicators $\mu _\pm = \exp \left( {\pm i\theta }
\right))$. We plot the lines of different bifurcations on a
parameter plane ($A_{01} ,\varphi _1 )$, analyzing equation
(\ref{eq19}) numerically. We note that negatively inclined
branches are unstable owing to tangent bifurcation. Thus lines
$\mu = 1$ in a parameter plane limit the regions of
multi-stability and hysteresis.

Results of numerical solution (\ref{eq19}) are performed in Fig.3.
Solid lines denote first doubling bifurcations, interrupted lines
correspond to tangent bifurcations, and dotted lines -- to Neimark
bifurcation. Vertical lines correspond to directions along which
gain characteristics, presented in Fig.2, were plotted. Letters
A,B,\ldots H mark rough transitions between different branches. As
seen from Fig.3 they coincide with hysteresis region, i.e. tangent
bifurcation. Varying of the second signal phase $\varphi _2 $ does
not influence dramatically the dynamics of the system, therefore
we restrict our numerical analysis to the case when $\varphi _2 =
0$. All results are periodic in $\varphi _1 $ with period equal to
$2\pi $.

Fig. 3a corresponds to Ikeda map. One observes the borders of
hysteresis region and doubling bifurcation line. When amplitude of
the second signal is rather small, there are no noticeable changes
in a parameter plane. The last can be observed only when $A_{02} $
increases to 0.4. There appear the regions of quasi-periodic
motion inside the region, limited by Neimark bifurcation lines
(fig 3c). With the increase of $I_{02} $ they grow in size, and
there exist two those regions, dwelling in different sheets of a
parameter plane (fig 3d).

\newpage
\section{Results of Numerical Modelling and Discussion}
\label{sec:results}

To investigate nonlinear dynamics of the coupled Ikeda maps
(\ref{eq15}) we performed numerical simulation in a wide range of
parameters. Our primary goal is to demonstrate that adding the
second signal component could cause qualitative changes in the
dynamics of the system in comparison to the single-frequency
driving. In Fig.~4 the typical results are presented as
bifurcation charts (phase diagrams) in the plane $\left( {A_{01}
,\varphi _1 } \right)$ for different values of $A_{02} $. The
choice of the parameters $A_{01} $ and $\varphi _1 $ (that have
meaning of external driving magnitude and frequency, respectively)
is quite natural for non-autonomous nonlinear oscillator. The
pictures are periodic in $\varphi _1 $ with the period $2\pi $.
The phase of the second signal $\varphi _2 $ does not strongly
affect the dynamics of the system and leads mainly to shift of the
bifurcation lines along the $\varphi _1 $ axes. Henceforth all the
presented results correspond to the case $\varphi _2 = 0$.

In Fig.~4 the regions of periodic oscillations with different
periods, quasi-periodic oscillations, and chaotic ones are shown.
In each point on the parameter plane we determine the type of
dynamical regime after sufficiently long transient process. For
that purpose we use phase portraits, power spectra, and spectra of
Lyapunov exponents. In the regions of periodic motion all the
exponents are negative, for quasi-periodic motion there are one
zero and three negative exponents, while in chaotic regimes one of
the exponents becomes positive \cite{Neomark,Landa,Ott}. We chose
the direction of scanning on the parameter plane to be
bottom{\-}up and from left to right and use the hereditary of
initial conditions. The last means when the transient process is
over and the dynamical regime in a point on the parameter plane is
determined we increase smoothly the value of the phase shift
$\varphi _1 $ taking the current values of variables as initial
conditions for a new point. When $\varphi _1 $ reaches the limit
$\varphi _1 = 2\pi $ we increase slightly the value of $A_{01} $
and return to $\varphi _1 = 0$. Note that owing to the bistability
the bifurcation charts are multi-folded, and only the parts
corresponding to the lower branch of the cavity transfer function
are shown in Fig.~4. To obtain the other parts of the charts one
should vary the scanning direction.

The bifurcation chart for $A_{02} = 0$ (Fig.~4(a)) coincide with
that for the single Ikeda map (\ref{eq1}), when built in
appropriate coordinates (see, for example \cite{Kuznetsov},
Figs.~3(b),4(b))\footnote{ In Ref.~\cite{Kuznetsov} the
bifurcation charts are built in the plane of parameters that
correspond to $A_0 $ and $1 \mathord{\left/ {\vphantom {1 \varphi
}} \right. \kern-\nulldelimiterspace} \varphi $ in the equation
(\ref{eq1}). After appropriate redrawing one can demonstrate that
the maps become $2\pi $-periodic in $\varphi $.}. In Fig.~4(a) one
can see the regions of period doublings and transition to chaos
via Feigenbaum scenario. We show only the first two doubling
bifurcations that are depicted with thin solid lines. The domains
of higher order bifurcations are not shown because of their small
sizes. In the region of chaotic motion there exist many domains of
periodic oscillations, which have a form known as
\textit{``crossroad area''} \cite{Ott,Kuznetsov}. The large ones
(with periods 3 and 4) are shown in Fig.~4(a). The exit from
crossroad areas occurs via Feigenbaum scenario.

For small enough amplitude of the second signal $A_{02} $
(Fig.~4(b)) the changes in the bifurcation chart compared with
Fig.~4(a) are hardly noticeable. Only slight deformation of the
borders of the oscillations of different types is observed.

With increasing of $A_{02} $ the bifurcation lines become more
complicated. Furthermore, in contrast to a single Ikeda map,
quasi-periodic motion can take place (Fig.~4(c,d)). The
quasi-periodic domains contain a large number of synchronization
tongues; the largest ones are shown in Fig.~4(c,d,e). Inside the
tongues the motion is periodic with rational winding numbers (the
values of the winding numbers are indicated in Fig.~4). We
emphasize that the winding number changes smoothly while moving in
the quasi-periodic domains\textbf{. }For example, in the domain
adjoining the region of periodic oscillations with period 1 in
Fig.~2(c), the winding number varies from 1 to $1 \mathord{\left/
{\vphantom {1 2}} \right. \kern-\nulldelimiterspace} 2$, and there
exist all synchronization tongues corresponding to the rational
winding numbers within the limits of variation. For sufficiently
large $A_{02} $ (Fig.~2(d,e)) quasi-periodic route to chaos
(Ruelle--Takens scenario) becomes possible.

It should be pointed out that variables $A_{1,2} $ are the
envelope amplitudes of spectral components with frequencies
$\omega _{1,2} $. Since a stable fixed point of the map
(\ref{eq15}) describes a steady-state solution when the amplitudes
are time-independent, $A_{1,2} = \mbox{const}$, the real physical
field $E\left( {x,t} \right)$ (\ref{eq6}) in general case (for the
frequencies $\omega _{1,2} $ supposed to be non-resonant, see
Sec.~\ref{sec:derivation}) exhibits quasi-periodic oscillations
with two independent frequencies, in contrast to the single Ikeda
map (\ref{eq1}) where steady-state solution corresponds to
periodic oscillations with frequency of external driving. In
non-stationary regimes when the amplitudes $A_{1,2} $ vary in
time, both components of the output signal become modulated. We
presume the carrier frequencies $\omega _{1,2} $ to be
sufficiently far off from each other ($\left| {\omega _1 - \omega
_2 } \right| \gg {2\pi } \mathord{\left/ {\vphantom {{2\pi } \tau
}} \right. \kern-\nulldelimiterspace} \tau $, $\tau $ is the time
delay), so their modulation spectra do not intersect. So, dealing
with an experimental device, it is possible to investigate the
dynamics of each component separately using a band-pass filter
tuned on one of the carrier frequencies.

When the first period doubling bifurcation occurs (appearance of a
2-period cycle from a stable point), the amplitudes exhibits
periodic self-pulsations with period $2\tau $. Therefore, this
bifurcation corresponds to the oscillatory Ikeda instability
caused by the delayed feedback
\cite{Neimark,Landa,Ikeda1,Ikeda2,Ikeda3}. The spectrum contains
side-bands (satellites) with frequencies $\omega _{1,2} \pm
n\omega _I $, $n = 1,2,\ldots $, where $\omega _I = \pi
\mathord{\left/ {\vphantom {\pi \tau }} \right.
\kern-\nulldelimiterspace} \tau $ --- the self-modulation
frequency. During transition to chaos the spectrum is enriched by
new sub-harmonic components with frequencies $\omega _{1,2} \pm
n{\omega _I } \mathord{\left/ {\vphantom {{\omega _I } {2^m}}}
\right. \kern-\nulldelimiterspace} {2^m}$.

\begin{figure}[t]
\leavevmode \centering
\includegraphics[width=0.7\columnwidth]{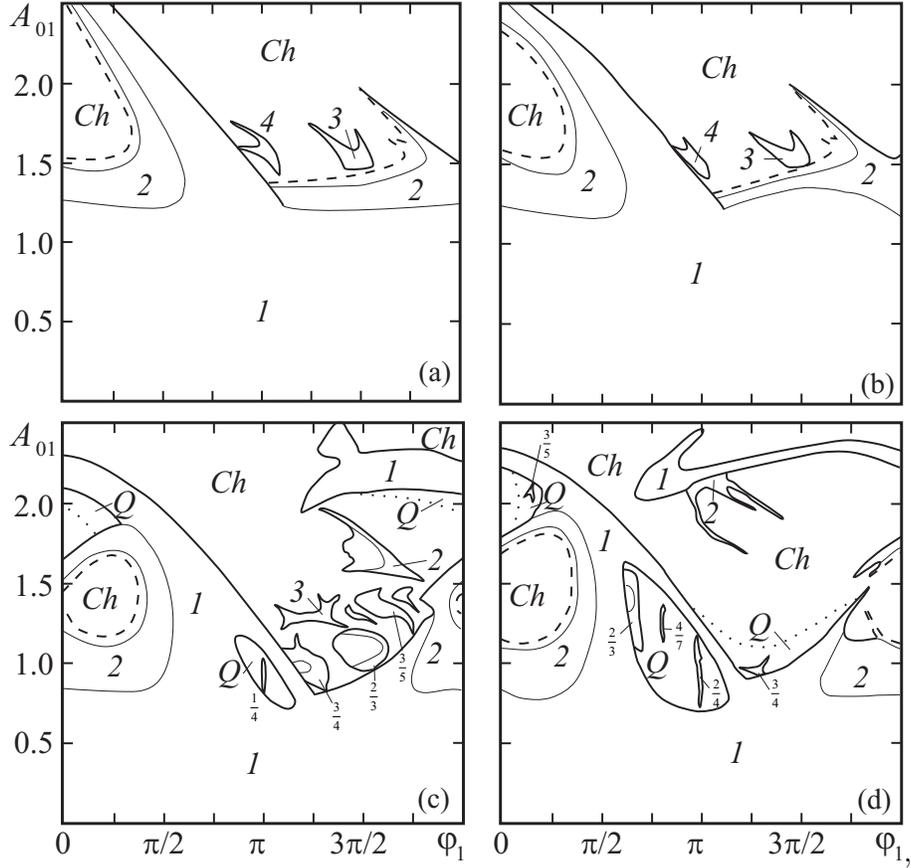},
\label{fig4} \caption{\small  Bifurcation charts on the plane of
parameters $\left( {A_{01} ,\varphi _1 } \right)$ for $\rho =
0.5$, $\varphi _2 = 0$, and $A_{02} = 0.0$ (a), 0.2 (b), 0.6 (c),
0.8 (d). Transitions to chaos via Feigenbaum scenario are shown by
dashed lines; quasi-periodic transitions to chaos are shown by
dotted lines; period doubling bifurcations are shown by thin solid
lines Only first two bifurcations are depicted, higher-order ones
are not shown for the lack of space. $Ch$ --- regions of chaotic
oscillations$\;\;Q$ --- regions of quasi-periodic oscillations,
numbers mark regions of periodic oscillations with corresponding
period. For synchronization tongues the values of winding number
are indicated.}
\end{figure}

On the contrary, in the quasi-periodic regime a fixed point loses
its stability via the Hopf bifurcation, when a unit circle is
crossed by a pair of complex conjugate multipliers and an
attractor is invariant curve \cite{Neimark,Landa,Ott}. The
side-band frequencies are $\omega _{1,2} \pm n\omega _Q $ with
$\omega _Q \ne \pi \mathord{\left/ {\vphantom {\pi {m\tau }}}
\right. \kern-\nulldelimiterspace} {m\tau }$. For the dynamics of
the total field $E\left( {x,t} \right)$ (\ref{eq6}) it corresponds
to periodic amplitude self-pulsations with period different from
$2\tau $, and the attractor is a 3D torus with three independent
frequencies (two career frequencies $\omega _{1,2} $ and that of
the self-pulsations, $\omega _Q )$. Inside the synchronization
tongues $\omega _Q $ is commensurable to the Ikeda frequency
$\omega _I $, i.e. a winding number $w = {\omega _Q }
\mathord{\left/ {\vphantom {{\omega _Q } {\omega _I }}} \right.
\kern-\nulldelimiterspace} {\omega _I }$ is rational, $w = p
\mathord{\left/ {\vphantom {p q}} \right.
\kern-\nulldelimiterspace} q$, $p$ and $q$ --- integer. The single
Ikeda map (\ref{eq1}) does not exhibit the Hopf bifurcation and
quasi-periodic motion
\cite{Neimark,Landa,Ikeda1,Ikeda2,Ikeda3,Moloney,Ott,Kuznetsov}.
Hence, we obtain a new type of instability caused by nonlinear
interaction of two driving signal components, i.e. by phase
cross-modulation. Note that the effects of phase cross-modulation
are very important in nonlinear optics, particularly for wave
propagation in optical fibers \cite{Agrawal}.

We point out that the results of direct computer modelling of
(\ref{eq1}) are of great agreement with the analysis of
characteristic equations undertaken in Sec.~\ref{sec:stat}. Some
distinctions are connected with the choice of scanning direction.

We observe three different routes to chaos: Feigenbaum period
doubling scenario, quasi-periodic (Ruelle--Takens) scenario, and
hard transition. Accordingly, three types of chaotic attractors
exist (Fig.~5). We use the following notation: $C_F $ for the
Feigenbaum attractor originated from period doubling bifurcation
sequence, $C_Q $ for the one formed from quasi-periodic motion,
and $C_I $ for the well-known Ikeda attractor \cite{Ikeda3} that
exists for high pumping levels and corresponds to fully-developed,
strongly irregular chaotic motion. The three described types
differ not only in visual appearance, but also in signature of
Lyapunov exponents: $C_Q $-type attractor has one positive, one
zero, and two negative exponents, while $C_F $ and $C_I $ types
have one positive and three negative exponents. No hyperchaotic
regimes with two positive exponents occur. We emphasize that for
the single Ikeda map (\ref{eq1}) only $C_F $ and $C_I $ types
exist and quasi-periodic route to chaos never takes place. We
represent in Fig.5 Feigenbaum-type attractor (a), $C_F $ ($\varphi
_1 = \pi \mathord{\left/ {\vphantom {\pi 6}} \right.
\kern-\nulldelimiterspace} 6$, $A_{01} = 1.3)$; (b) attractor
originated from the quasi-periodic route to chaos, $C_Q $
($\varphi _1 = {3\pi } \mathord{\left/ {\vphantom {{3\pi } 2}}
\right. \kern-\nulldelimiterspace} 2$, $A_{01} = 1.5)$; (c)
Ikeda-type attractor, $C_I $ ($\varphi _1 = \pi $, $A_{01} =
2.5)$. Other parameters are the same as for Fig.~4(c).

\begin{figure}[h]
\leavevmode \centering
\includegraphics[width=0.9\columnwidth]{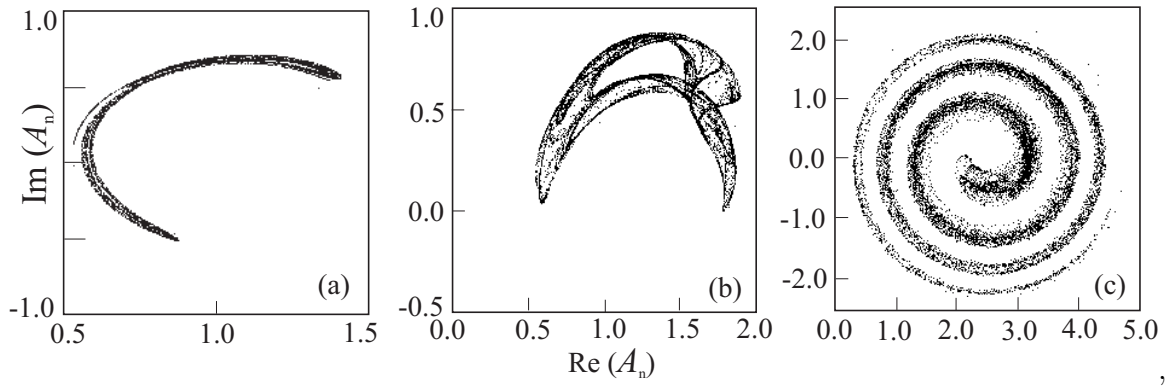},
\label{fig5} \caption{Chaotic attractors of different types}
\end{figure}

From Fig.~3 one can see that the bifurcation charts are very
complicated, and thus, a great variety of bifurcation sequences
could be observed while moving in different ways on the parameter
plane. For better understanding the behavior of the system we
built single-parameter bifurcation diagrams: $Re\left( {A_n^1 }
\right)$ plots versus $A_{01} $. Some examples are presented in
Fig.~6 for different values of $\varphi _1 $. All diagrams are
plotted with smooth increase of $A_{01} $ with hereditary of
initial conditions for $\varphi _1 = \pi \mathord{\left/
{\vphantom {\pi 6}} \right. \kern-\nulldelimiterspace} 6$ (a),
$\varphi _1 = \pi $ (b), and $\varphi _1 = {3\pi } \mathord{\left/
{\vphantom {{3\pi } 2}} \right. \kern-\nulldelimiterspace} 2$ (c).
Other parameters are the same as for Fig.~4(c).

\begin{figure}[h]
\leavevmode \centering
\includegraphics[width=0.37\columnwidth]{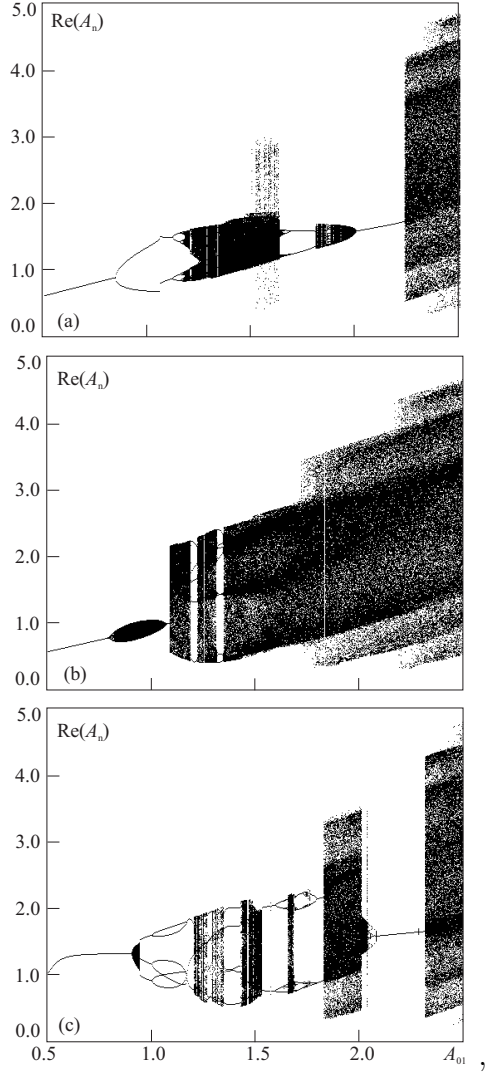},
\label{fig6}

\caption{\small Single-parameter bifurcation diagrams $Re\left(
{A_n^1 } \right)$ versus $A_{01} $}
\end{figure}

The first diagram (Fig.~6(a)) shows period doubling bifurcations
and transition to chaos via Fiegenbaum scenario at $A_{01} \approx
1.2$. In the region of chaotic motion several windows of
periodicity are clearly seen. At $A_{01} \approx 1.5$ the $C_F
$-type attractor begins to transform in the $C_I $-type one. When
$A_{01} \approx 1.63$ one observes a hard transition between
different types of chaotic oscillations known as a crisis of
chaotic attractors. In that case chaotic attractor abruptly
expands or contrarily shrinks, that is explained by its collision
with non-stable periodic orbit \cite{Grebogi,Ott}. Further
increase of $A_{01} $ leads to the reverse doubling bifurcation
sequence. At $A_{01} \approx 1.8$ a hard transition to
quasi-periodic motion takes place. Inside the region of
quasi-periodicity one can see several windows of synchronization.
With increase of $A_{01} $ the quasi-periodic attractor shrinks,
and at $A_{01} \approx 2.0$ the motion again becomes periodic.
Finally, the hard transition to Ikeda attractor $C_I $ takes place
at $A_{01} \approx 2.25$.

On the second diagram (Fig.~6(b)) one can see a region of
quasi-periodic oscillations that appears and disappears softly
(i.e. via direct and reverse Hopf bifurcations) at $A_{01} \approx
0.8$ and $A_{01} \approx 1.1$, respectively. Hard transition to
chaotic attractor $C_I $ occurs at $A_{01} \approx 1.25$.
Subsequent increase of $A_{01} $ causes the expansion of the
attractor $C_I $.

The bifurcation diagram presented on Fig.~4(c) illustrates soft appearance
of quasi-periodic oscillations ($A_{01} \approx 0.9)$. A number of
synchronization windows with different periods can be observed, inside some
of them period doublings takes place. At $A_{01} \approx 2.0$ a crisis of
chaotic attractors is clearly visible. The bifurcation sequence ends with a
hard transition to Ikeda attractor at $A_{01} \approx 2.3$.

\section{Concluding remarks} \label{sec:concluding}

In this paper, we discussed an extension of the Ikeda system for
the case of multi-frequency external signal. The system of coupled
Ikeda maps was derived to describe the dynamics of slowly varying
amplitudes of spectral components. For the simplest case of
double-frequency driving results of numerical simulation were
presented, that confirm that the dynamics is much more diverse
than for the single-frequency driving. Various transitions to
chaos were observed: the Feigenbaum period doubling scenario,
quasi-periodic (Ruelle--Takens) scenario, hard transition to
chaos, and crisis of chaotic attractors. Applying of the second
input signal gives the opportunity to efficient control of the
dynamics of the system. Changing amplitude and phase of the second
signal component one can either suppress the chaotic oscillations
or conversely cause the steady-state oscillations to be unstable.
These results are significant for prospects to use nonlinear ring
cavities as logical elements in optical computers \cite{Rosanov}.

However, the validity of the model (\ref{eq15}) needs further
consideration, since a number of simplifying assumptions has been
made. First of all, we neglected transverse effects (diffraction,
self-focusing, etc.) that lead to a wide class of transversal
instabilities \cite{Rosanov,Moloney}. We also did not take into
account group velocity dispersion effects that may cause
modulation (Benjamin--Feir) instability and formation of solitons
\cite{Rosanov,Agrawal}. All these phenomena usually play an
important role in nonlinear optics, and their consideration will
be a subject of further investigation.

Finally, we note that Ikeda map describes non only a nonlinear
ring cavity, but also many other interesting physical systems,
e.g. a periodically kicked nonlinear oscillator \cite{Kuznetsov}.
Evidently, the system of coupled maps (\ref{eq15}) can be derived
for a quasi-periodically kicked oscillator.

\section{Acknowledgements} \label{sec:ack}

This work was supported by USA Civilian Research and Development Foundation
(Award No REC-006) and Russian Foundation for Basic Research (grants No.
02-02-16351 and 02-02-06315).

\end{document}